\documentclass[epj]{svjour}
\usepackage{latexsym}
\usepackage{graphicx}

\newcommand{\mev}{\textrm{ MeV}}
\newcommand{\MeV}{\textrm{ MeV}}

\newcommand{\Sgs}{\Sigma^*}

\newcommand{\bk}{\bar{K}}

\newcommand{\be}{\begin{equation}}
\newcommand{\ee}{\end{equation}}
\newcommand{\ba}{\begin{eqnarray}}
\newcommand{\ea}{\end{eqnarray}}
\newcommand{\Ls}{{\Lambda(1520)}}
\newcommand{\rw}{\rightarrow}

\begin{document}

\title{\hfill{\small FZJ--IKP(TH)--2006--01}
\\
Testing the nature of the 
{\boldmath$\Lambda(1520)$} resonance in proton-induced
production}

\author{
L.~Roca\inst{1}, C.~Hanhart\inst{2}, 
E.~Oset\inst{3}, Ulf-G. Mei{\ss}ner\inst{4,2}}

\institute{
Departamento de F\'{\i}sica, Universidad de Murcia, 
E-30100 Murcia, Spain
\and
Institut f\"ur Kernphysik (Theorie), Forschungszentrum J\"ulich,
D-52425 J\"ulich, Germany
\and
Departamento de F\'{\i}sica Te\'orica and IFIC,
Centro Mixto Universidad de Valencia-CSIC,
Institutos de
Investigaci\'on de Paterna, Aptdo. 22085, 46071 Valencia, Spain
\and
HISKP (Th), Universit\"at Bonn, Nu{\ss}allee 14-16, D-53115 Bonn, Germany
}

\date{Received:  / Revised version: }
%

\abstract{ The $\Lambda(1520)$ resonance has been recently
studied in a unitarized coupled channel formalism with 
$\pi\Sigma(1385)$, $K\Xi(1530)$, $\bar{K}N$ and $\pi\Sigma$ as
constituents blocks. We provide a theoretical study of the
predictions of this model  in physical
observables of the $pp\to pK^+K^-p$ and  $pp\to
pK^+\pi^0\pi^0\Lambda$ reactions. In particular, we show that
the ratio  between the $\pi^0\pi^0\Lambda$ and
$K^-p$ mass distributions can provide valuable information on
the ratio of the couplings of the $\Lambda(1520)$ resonance to
$\pi\Sigma(1385)$ and $\bar{K}N$ that the theory predicts.
Calculations are done for energies which are accessible in an
experimental facility like COSY at J\"ulich or the developing CSR 
facility at Lanzhou.
}

\authorrunning{L.~Roca et al.}

\titlerunning{Testing the nature of the {\boldmath$\Lambda(1520)$} 
resonance in proton-induced production}

\maketitle

\section{Introduction}

The low-lying negative parity resonances $J^P=1/2^-$, $3/2^-$, 
have recently attracted much attention as many of them can
qualify as dynamically generated resonances from the interaction
of mesons and baryons. In particular, much progress has been done
interpreting the low lying $1/2^-$ resonances as dynamically
generated from the interaction of the octet of pseudoscalar
mesons with the octet of stable baryons
\cite{Kaiser:1996js,Oset:1997it,Meissner:1999vr,Oller:2000fj,Garcia-Recio:2002td,Hyodo:2003jw}
which allows one to make predictions for resonance formation in
different reactions \cite{Oller:2000ma}. One of the surprises on
this issue was the realization that the $\Lambda(1405)$ resonance
is actually a superposition of two states, a wide one coupling
mostly to $\pi\Sigma$ and a narrow one coupling mostly to $\bar K
N$ \cite{Oller:2000fj,Jido:2003cb,Garcia-Recio:2003ks}. The
performance of a recent experiment on the
$K^-p\to\pi^0\pi^0\Sigma^0$ reaction \cite{Prakhov:2004an} and
comparison with older ones, particularly the $\pi^-p\to
K^0\pi\Sigma$ reaction \cite{Thomas:1973uh}, has brought evidence
on the existence of these two $\Lambda(1405)$ states
\cite{Magas:2005vu}. Extension of these works to the interaction
of the octet of pseudoscalar mesons with the decuplet of baryons
has led to the conclusion that the low lying $3/2^-$ baryons are
mostly dynamically generated objects \cite{lutz,Sarkar:2004jh}.
One of these states is the $\Ls$ resonance which was generated
from the interaction of the coupled channels $\pi\Sigma(1385)$
and $K\Xi(1530)$. From the experimental point of view this
resonance is of particular interest in searches of pentaquarks in
photononuclear reactions \cite{nakano,Hicks:2005gp} in $\gamma
p\to K^+K^-p$ and $\gamma d\to\ K^+K^-np$. Since getting signals
for pentaquarks involves cuts in the spectrum and subtraction of
backgrounds, the understanding of the resonance properties and
the strength of different reactions in the neighborhood of the
peak becomes important in view of a correct interpretation of
invariant mass spectra when cuts and background subtractions are
made. From the theoretical point of view, in the studies of 
Refs.~\cite{lutz,Sarkar:2004jh}, the $\Lambda(1520)$ is build up
from the  $\pi\Sigma(1385)$ and $K\Xi(1530)$ and couples mostly
to the first channel, to the point that, in this picture, the
state would qualify as a quasibound $\pi\Sigma^*$ state. Indeed,
the nominal mass of the $\Ls$ is a few$\MeV$ below the average of
the $\pi\Sigma^*$ mass. However, the PDG \cite{Eidelman:2004wy}
gives a width of $15\mev$ for the $\Ls$, with branching ratios of
$45$\%  into $\bar KN$ and $43$\% into $\pi\Sigma$, and only a
small branching ratio of the order of $4$\%  for $\pi\Sigma^*$
which could be of the order of $10$\% according to some analysis
\cite{Mast:1973gb} which claims that about $85$\%  of the decay
into $\pi\pi\Lambda$ is actually $\pi\Sigma^*$. The association
of $\pi\pi\Lambda$ to $\pi\Sigma^*$ in the peak of the $\Ls$ is a
non trivial test since one has no phase space for $\pi\Sigma^*$
excitation and only the width of the $\Sigma^*$ allows for this
decay, hence precluding the reconstruction of the $\Sigma^*$
resonant shape from the $\pi\Lambda$ decay product.  

One step
forward in the understanding of the $\Lambda(1520)$ resonance has
been possible thanks to a recent reaction
$K^-p\to\pi^0\pi^0\Lambda$, experimentally performed in
Ref.~\cite{Prakhov:2004ri} and theoretically studied in
Refs.~\cite{Sarkar:2005ap,we}. The reaction proceeds mostly via 
$K^-p\to\pi^0(\Sigma^{*0})\to\pi^0(\pi^0\Lambda)$.  This is seen
at energies of the $K^-$ such that
$\sqrt{s}>M_{\pi^0\Sigma^{*0}}$, where the reconstruction of the
$\pi^0\Lambda$ invariant mass produces the $\Sigma^{*0}$
resonance shape \cite{Prakhov:2004an,Sarkar:2005ap}. However, at
energies of the $K^-$ where $\sqrt{s}\simeq M_{\Ls}$, the
$\Sigma^{*0}$ is only produced through the tail of the resonance.
Yet, a formalism using explicitly the $\Sigma^{*0}$ propagator
allows us to study properly the $K^-p\to\pi^0\pi^0\Lambda$
reaction assuming the same $\pi^0\Sigma^{*0}$  as the dominant
mechanism.

On the other hand, the large branching ratios to $\bar K N$ and
$\pi \Sigma$, of the order of $90$\% together, indicate that the
$\bar K N$ and $\pi\Sigma$ channels must play a relevant role in
building up the $\Ls$ resonance. In the work of
Refs.~\cite{Sarkar:2005ap,we} this problem was tackled  by
performing a coupled channel analysis of the $\Ls$ data with
$\pi\Sigma^*$, $K\Xi^*$, $\bar{K}N$ and $\pi\Sigma$, the first
two channels interacting in $s$-wave and the last two channels in
$d$-wave to match  the $3/2^-$ spin and parity of the $\Ls$
resonance. In Ref.~\cite{we} it was also shown that although the
$\pi\Sigma^*$ remains with the largest coupling to $\Ls$, its
strength is reduced with respect to the simpler picture of only
$\pi\Sigma^*$ building up the resonance, and at the same time
there is a substantial coupling to $\bar KN$ and $\pi\Sigma$
which distorts the original $\pi\Sigma^*$ quasibound picture and
makes the $\bar KN$ and $\pi\Sigma$ channels relevant in the
interpretation of different reactions.

In the present work we propose two reactions to test the
non-trivial predictions of the unitarized coupled channel model
of Ref.~\cite{we} regarding the couplings of the $\Ls$ resonance
to the different channels. These reactions are $pp\to pK^+K^-p$
and  $pp\to pK^+\pi^0\pi^0\Lambda$ close to the $\Ls$ threshold.
Particularly, we show that a measurement of the ratio  between
the $\pi^0\pi^0\Lambda$ and $K^-p$ mass distributions for $pp\to
pK^+\pi^0\pi^0\Lambda$ and $pp\to pK^+K^-p$ reactions
respectively is an excellent model independent  test of the ratio
between the $\Ls$ coupling to $\pi\Sigma(1385)$ and to
$\bar{K}N$.

The structure of the paper is as follows. In section
\ref{sec:model} the unitarized coupled channel
model of Ref.~\cite{we} is summarized. In section~\ref{sec:pprod}
the dominant mechanism to produce the $\Ls$ in p-induced
reactions and the implementation of the unitarized coupled
channel model into it is described. Finally, in
section~\ref{sec:results} we show the results for the mass
distributions in the two reactions studied.

\section{Summary of the unitarized coupled channel formalism}
\label{sec:model}

\begin{figure}
\begin{center}
\includegraphics[width=0.48\textwidth,angle=0]{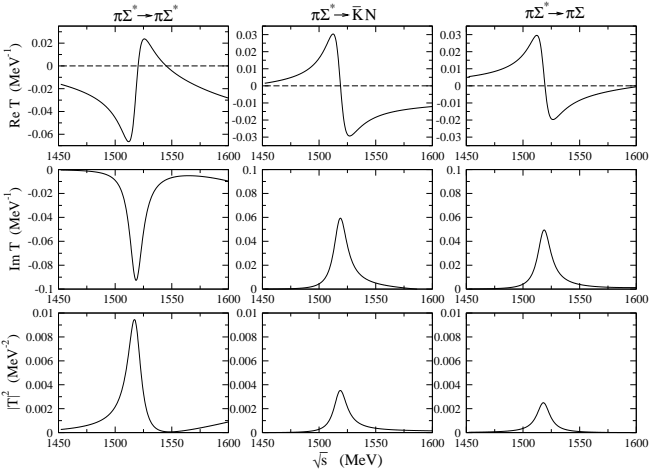}
\end{center}
\caption{Unitary amplitudes involving the $\pi\Sigma^*$ channel.
 From left to right:
 $\pi\Sigma^*\to\pi\Sigma^*$, $\pi\Sigma^*\to \bar K N$ and 
$\pi\Sigma^*\to \pi\Sigma$.}
\label{fig:Tij}
\end{figure}

In Ref.~\cite{we} the $\Lambda(1520)$ resonance was
studied within a coupled channel formalism including the
$\pi\Sigma^*$, $K\Xi^*$ in $s$-wave and the $\bar K N$  and
$\pi\Sigma$ in $d$-waves.
 Unitarity was implemented by means of 
  the Bethe-Salpeter (BS) equation
in the evaluation of the different scattering
amplitudes, which reads
\be
T=V+VGT \Rightarrow T=[1-VG]^{-1}V.
\label{eq:bethe}
\ee
\noindent
The kernel of the BS equation is given by the following
transition potentials \cite{we}
\renewcommand{\arraystretch}{1.25}
\be
V=\left| 
\begin{array}{cccc}
C_{11}(k_1^0+k_1^0)\ & C_{12}(k_1^0+k_2^0) & \gamma_{13}\,q_3^2 &\gamma_{14}\,q^2_4 \\
C_{21}(k_2^0+k_1^0)\ & C_{22}(k_2^0+k_2^0) & 0 & 0 \\
\gamma_{13}\,q_3^2 & 0 & \gamma_{33}\, q^4_3 & \gamma_{34} \,q_3^2 \,q^2_4\\
\gamma_{14}\,q^2_4  & 0 & \gamma_{34} \,q_3^2 \,q_4^2 &  \gamma_{44}\, q^4_4
\end{array}
\right|~,
\label{eq:Vmatrix}
\ee
\noindent
where the elements  $1$, $2$, $3$ and $4$ denote 
$\pi\Sigma^*$, $K\Xi^*$,  $\bar K N$  and
$\pi\Sigma$ channels respectively.
In Eq.~(\ref{eq:Vmatrix}),
\be
q_i=\frac{1}{2\sqrt{s}}\sqrt{[s-(M_i+m_i)^2][s-(M_i-m_i)^2]},
\ee
$k_i^0=\frac{s-M_i^2+m_i^2}{2\sqrt{s}}$ 
and $M_i(m_i)$ is the baryon(meson) mass. 
The coefficients $C_{ij}$ are $C_{11}=\frac{-1}{f^2}$,
 $C_{21}=C_{12}=\frac{\sqrt{6}}{4f^2}$ and $C_{22}=\frac{-3}{4f^2}$,
where $f$ is $1.15f_\pi$, with $f_\pi$ ($=93\mev$) the pion decay constant,
which is an average between $f_\pi$ and
  $f_K$ as was used in Ref.~\cite{Oset:1997it} in
  the related problem of the dynamical 
  generation of the $\Lambda(1405)$.
In Eq.~(\ref{eq:bethe}) $G$ stands for a diagonal matrix containing
the loop functions involving a baryon  and a meson 
which are regularized by means of two subtraction constants
\cite{we}:
one for the s-wave channels ($a_0$) and another one
 for the d-wave channels ($a_2$).
 The treatment of Eqs.~(1) in Refs.~\cite{Sarkar:2005ap,we} relies upon a
 dispersion relation on $T^{-1}$ which allows the on-shell factorization
 out of the loops of the kernel, $V$, of the Bethe-Salpeter equation
 \cite{Oset:1997it,ollerND}.
The matrix 
elements $V_{11}$, $V_{12}$, $V_{21}$ and $V_{22}$ come from
 the lowest order chiral Lagrangian involving the decuplet of baryons and
the octet of pseudoscalar mesons, as discussed in
 Ref.~\cite{manohar,lutz,Sarkar:2004jh}.
The 
unknown parameters in the $V-$matrix, as well as the
subtraction constants of the loop functions, were obtained
by a fit to $\bar K N\to\bar K N$ and $\bar K N\to \pi\Sigma$
partial wave amplitudes.
The values obtained are shown in Table~\ref{tab:fit}~\cite{we}.
\begin{table}
\caption{Loop subtraction constants and parameters
 of the potentials. The $\gamma_{13}$ and $\gamma_{14}$
 are given in units of GeV$^{-3}$; and $\gamma_{33}$,
 $\gamma_{44}$ and $\gamma_{34}$ in units of GeV$^{-5}$.}
\label{tab:fit}     
\begin{center}
\begin{tabular}{ccccccc}
\hline\noalign{\smallskip}
$a_0$ & $a_2$ & $\gamma_{13}$ & $\gamma_{14}$&
    $\gamma_{33}$ &  $\gamma_{44}$ &  $\gamma_{34}$   \\
\noalign{\smallskip}\hline\noalign{\smallskip}
 $-1.8$ & $-8.1$ & $98$ & $110$ & $-1730$
 & $-730$ & $-1108$ \\
\noalign{\smallskip}\hline
\end{tabular}
\end{center}
\end{table}
Despite the apparent large number of parameters, it is worth mentioning that
the dominant potentials are $V_{11}$, $V_{12}$, $V_{21}$ and $V_{22}$ which
have no freedom. 
Let us recall that we do not include $\pi \Sigma^*$
 amplitudes in our fit. Hence, the amplitudes involving
  this channel are  genuine predictions of the theory 
  and are shown in Fig.~\ref{fig:Tij}.

From the scattering amplitudes we can also extract the effective couplings  of
the $\Ls$ to all the different channels, which can be
obtained via an analytic continuation of the amplitude in the complex plane.
Near the pole we may write up to regular terms
\be
T_{ij}(\sqrt{s})=\frac{g_i g_j}{\sqrt{s}-M_\Ls + i \Gamma_\Ls /2}
\label{eq:Tcoup}
\ee 
from where we have
 \be
g_i g_j=-\frac{\Gamma_\Ls}{2} \frac{|T_{ij}(M_\Ls)|^2}
{Im[T_{ij}(M_\Ls)]},
\ee
\noindent 
where $M_\Ls$ is the position of the peak in $|T_{ij}|^2$ and 
$\Gamma_\Ls=15.6\mev$.
The couplings obtained are shown in Table~\ref{tab:coup}.

\begin{table}
\caption{Couplings of the $\Ls$ resonance
 to the different channels}
\label{tab:coup}     
\begin{center}
\begin{tabular}{cccc}
\hline\noalign{\smallskip}
$g_1$ &$g_2$ &$g_3$ &$g_4$  \\
\noalign{\smallskip}\hline\noalign{\smallskip}
 $0.91$ & $-0.29$ & $-0.54$ & $-0.45$ \\
\noalign{\smallskip}\hline
\end{tabular}
\end{center}
\end{table}

\section{Proton-induced $\Ls$ production}
\label{sec:pprod}

The amplitudes involving $\pi\Sigma^*$
channels can be investigated in particular in those reactions where this
channel plays a significant role. In Ref.~\cite{we}, these
amplitudes were checked in the $K^-p\to\Lambda\pi\pi$,  $\gamma
p\to K^+K^-p$, $\gamma p\to K^+\pi^0\pi^0\Lambda$ and $\pi^-
p\to K^0 K^-p$ reactions, leading to a good reproduction to the
experimental results.

In the present work we  apply the model to the 
$pp\to pK^+(\Ls)\to pK^+(\pi^0\pi^0\Lambda)$ and 
$pp\to pK^+(\Ls)\to pK^+(K^-p)$ reactions at energies slightly above the $\Ls$
production threshold, which are attainable at a facility like
COSY \cite{meier,Zychor:2005sj}.

As explained in  Appendix~\ref{app:ampl}, only a single partial wave, with the
initial $pp$ pair in the
$^1D_2$\footnote{Here we use the standard notation for the $NN$ partial waves:
$^{2S+1}L_J$, with $S$, $L$, $J$ for the total spin, the angular momentum and
  the total angular momentum.},
contributes to the process $pp\to K^+ p\Ls$ near threshold under the
assumption that the $\Lambda (1520)$KN system in the intermediate state is in
an $s$--wave in all subsystems. 
 In addition, since the production
is characterized by a large momentum transfer, we can assume the production
operator to be (largely) independent of the final momenta, that are
constrained to small values because of the chosen kinematics. Thus the only
significant source of an energy or momentum dependence with respect to the final
particles should be their various final state interactions. Therefore, to
deduce information on the final state interactions in large momentum transfer
reactions, no detailed knowledge of the production operator is necessary (for
a recent review on these issues we refer to Ref. \cite{report} and the
references therein). 

There are various ways to derive the relevant matrix elements. The one most
commonly used is based on the irreducible tensor techniques and the
corresponding formulas are given in Appendix B. Here on the main text, on the
other hand, we will describe a method for the construction of matrix elements
that is more transparent and allows us to easier explain the relevant
physics. Needless to say that the final results in the two approaches are
identical. 

According to the selection rules given above the transition $pp\to
\Lambda(1520) Kp$ can be parameterized by a single constant $C$ and a fixed
operator structure which reads for non--relativistic final states
\begin{equation}
W=C\left((\vec u_\Ls^\dagger \cdot \vec p)(\vec \sigma \cdot \vec p)
\sigma_2 u_p^*\right)\phi_K\left(u_p^T\sigma_2u_p\right) \ ,
\label{wdef}
\end{equation}
where $\vec u_\Lambda$ and $u_p$ denote the spinors for the $\Lambda(1520)$ as well as
the nucleons. Note, by construction $\vec \sigma \cdot \vec u_\Ls$
vanishes. It is this identity that ensures that it is only the $pp$ $D$ wave
that contributes to the above transition operator. For the initial momentum
we use $\vec p$, and $\vec \sigma$
denotes the standard three vector of Pauli matrices. In the expressions below
we will omit the spinors from the initial state to simplify notations.

\begin{figure}
\begin{center}
\includegraphics[width=0.2\textwidth,angle=0]{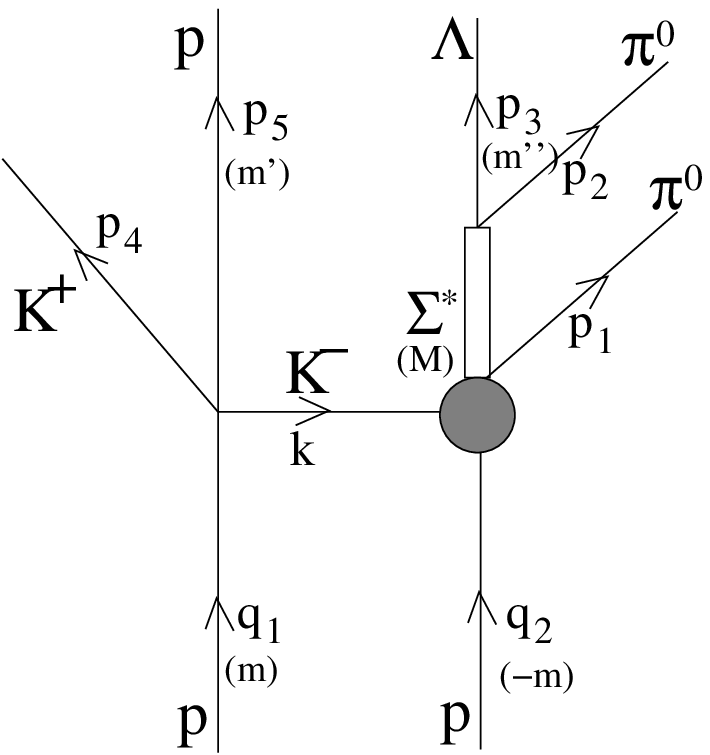}
\end{center}
\caption{Mechanism considered for 
$pp\to pK^+(\Ls)\to pK^+(\pi^0\pi^0\Lambda)$}
\label{fig:diagrampipiL}
\end{figure}

To come to the corresponding expressions for the full transitions---including
the decay of the $\Ls$---we need to contract the transition operator $W$ with
further operators, namely the spin transition operator $\vec S$
normalized as 
$$
\vec S_i\vec S^\dagger_j=\frac13\left(2\delta_{ij}-i\epsilon_{ijk}\sigma_k\right) \
, 
$$
as it
occurs in the non--relativistic version of the Rarita--Schwinger propagator, 
as well as the relevant vertices and propagators.
As a result for the decay matrix element for the chain
$$pp\to K^+p\Ls \to K^+p(\pi^0\Sigma^{*\, 0})\to K^+p(\pi^0[\Lambda \pi^0])$$
we need to replace $\vec u_{\Ls \, i}^\dagger$ in Eq. (\ref{wdef}) by
\begin{eqnarray} 
\nonumber \Gamma_{\pi\pi}&=&-\frac1{\sqrt{3}}
g_1\frac{f_{\Sigma^*\pi\Lambda}}{m_\pi}\bar u_\Lambda
(p_3)(\vec S \cdot \vec p_1) \vec S^\dagger_i
G_{\Lambda^*}(s_{\Lambda^*}) G_{\Sigma^*}(s_{\Sigma^{*}}^1) \ ,
\end{eqnarray}
where we used for the transition $\Ls\to \Sigma^*\pi$ a constant matrix
element and for $\Sigma^*\to \Lambda \pi$ the standard vertex $\vec S \cdot
\vec p_1$. In the expression $-g_1/\sqrt{3}$ is the effective coupling for
$\Lambda^*\to \pi^0\Sigma^{*\, 0}$ including the isospin factor.  Here
$G_{\Lambda^*}$ denotes the propagator of the $\Ls$,
implicit in the dashed circle in Fig.~\ref{fig:diagrampipiL},
 which is a function of
$s_{\Lambda^*}=(p_1+p_2+p_3)^2$.  $G_{\Sigma^*}$ is the propagator of the
$\Sigma^*$ being a function of $s_{\Sigma^{*}}^1=(p_2+p_3)^2$, where we use
the standard Flatte--parametrization
$$
G_{\Sigma^*}(s_{\Sigma^*})=\frac{1}
{\sqrt{s_{\Sgs}}-M_{\Sgs}+i\Gamma_{\Sgs}(\sqrt{s_{\Sgs}})/2}
$$
Note, to make contact with the $T$--matrices derived in the previous section we
need to replace $g_1G_{\Lambda^*}$ by the corresponding channel $T$ matrix. We come
back to this below. 

After some algebra we  get for the complete
matrix element in the $\pi\pi\Lambda$ channel
\begin{eqnarray}\nonumber
A^{\pi\pi\Lambda}(p_1,p_2)&=&
-\frac{Cg_1}{\sqrt{3}}\frac{f_{\Sigma^*\pi\Lambda}}{m_\pi}
G_{\Lambda^*}(s_{\Lambda^*})\bar u_\Lambda (p_3)\vec \sigma_i\sigma_2
u^*(p_5) \\
& &\!\!\!\!\!\!\!\!\!\!\!
\times\left\{{\cal P}(p_1)_iG_{\Sigma^*}(s_{\Sigma^{*}}^1)
+{\cal P}(p_2)_iG_{\Sigma^*}(s_{\Sigma^{*}}^2)\right\} \, ,
\end{eqnarray}
where ${\cal P}(k)_i=(\vec k\cdot \vec p)p_i-\frac13p^2\vec k_i$ projects on the
initial $D$--wave, as already explained above.
One observes that the $\Lambda p$ system in the final state occurs solely in
the spin triplet in line with the findings of Appendix B.

On the other hand, for the reaction chain
$$pp\to K^+p\Ls \to K^+p(K^-p)$$
we need to  replace $\vec u_{\Ls \, i}^\dagger$ in Eq. (\ref{wdef}) by
\begin{equation}
\Gamma_{K}=-\frac1{\sqrt{2}}g_3\bar u(p_1)
(\vec \sigma \cdot \vec p_3)(\vec S \cdot \vec p_3)\vec
S^\dagger_i \ . 
\end{equation}
 We get after some standard
manipulations
\begin{eqnarray}\nonumber
A^{Kp}(p_1,p_2)&=&-\frac1{\sqrt{2}}
Cg_3\left\{ \bar u (p_1)\sigma_2u(p_2)({\cal P}(p_3)\cdot p_3)\right. \\
\nonumber
& & \qquad \qquad \qquad \times \left. \left(
G_{\Lambda^*}(s_1)+G_{\Lambda^*}(s_2)\right)\right. \\ \nonumber
& & \quad +  i\bar u (p_1)\sigma_k\sigma_2u(p_2)(\vec p_3\times \vec
  p)_k(\vec p_3\cdot \vec p) \\
& & \qquad \qquad \qquad \times \left. \left(
G_{\Lambda^*}(s_1)-G_{\Lambda^*}(s_2)\right)\right\} \ .
\end{eqnarray}
Here the first term contains the two proton system in a spin singlet state
(e.g. $^1S_0$), whereas the second term contains the $pp$ system in a spin
triplet state.  The second term implies $L=odd$ for the two protons. However,
as explained in Appendix~\ref{app:ampl}, the energies of the proton are such
that only $L=0$ is relevant.  In addition,
$G_{\Lambda^*}(s_1)-G_{\Lambda^*}(s_2)$ is much smaller than the sum.
Therefore we will only include the first term in what follows.

\begin{figure}
\begin{center}
\includegraphics[width=0.2\textwidth,angle=0]{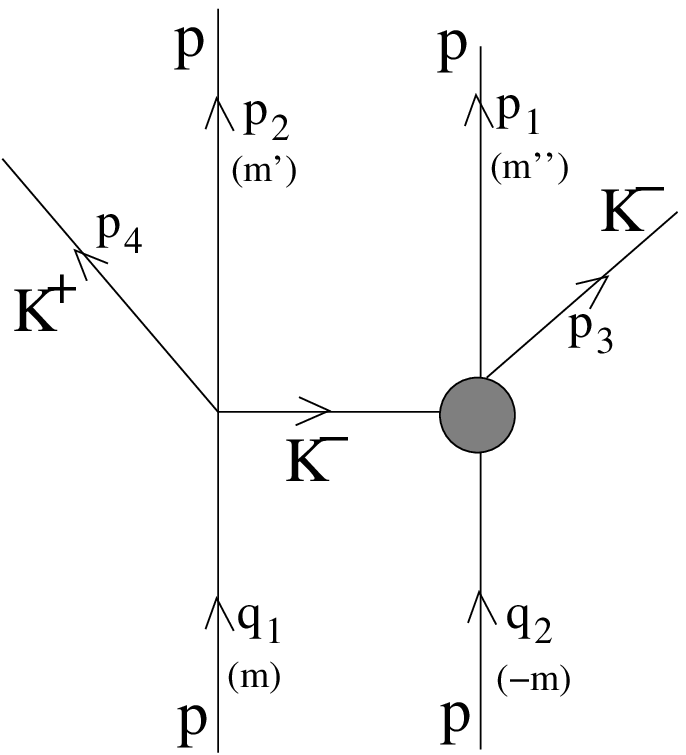}
\end{center}
\caption{Mechanism considered for 
$pp\to pK^+(\Ls)\to pK^+(K^-p)$}
\label{fig:diagramKp}
\end{figure}

As long as we study ratios only, the only unknown in the above
expressions, $C$, drops out and all distributions turn out to be predictions
of the model. However, to make contact between the coupled channel T--matrices of the
previous section and the expressions just derived we use a particular
production mechanism as shown in Figs. \ref{fig:diagrampipiL} and \ref{fig:diagramKp}.
Thus the transition operator parameterized as $C$  contains---in this
model---the $KK\bar NN$ vertex, where the outgoing $K^+$ is produced, as well
as a $\bar K p$ vertex on the other nucleon. We may thus pull the latter out
of the definition of $C$ and write
$$
-\frac1{\sqrt{2}}\tilde C g_3 = C \ .
$$
With this definition it is straight forward to show that
\begin{eqnarray}
\tilde C &=&\frac{1}{2f^2}(p_4^0-k^0)
\frac{1}{k^2-m_k^2} \ 
\end{eqnarray}
for the transition operator and
\begin{eqnarray}
\left(g_1\right)G_\Ls \left(\frac{g_3 p^2}{\sqrt{3}}\right)
&=& T_{\bar{K}N \to\pi\Sigma^*} \
, \\
\left(\frac{g_3 p_3^2}{\sqrt{3}}\right)G_\Ls\left(\frac{g_3 p^2}{\sqrt{3}}\right)
&=& T_{\bar{K}N \to\bar{K}N} \ 
\end{eqnarray}
for the meson--baryon $T$--matrices.
This completes the evaluation of the production matrix elements.

By the described method the antisymmetrization of the two proton states
 has already been taken into account since only
allowed partial wave were considered. The
symmetrization of the two identical $\pi^0$ has to be considered by adding the
contribution obtained by changing $p_1\leftrightarrow p_2$ (see details in
Appendix~\ref{app:ampl}). Hence, $|Amp|^2$ to be used in the cross section
(c.f. Eq.~(\ref{eq:crosspipiL}) in the appendix) is
given by

\be
|Amp|^2=\sum_{S_z}|A^{\pi\pi\Lambda}(p_1,p_2)+A^{\pi\pi\Lambda}(p_2,p_1)|^2~.
\ee
For the calculation of the cross sections one has to evaluate 
a four-- and five--body phase
space. The method we used is given in detail in Appendix~\ref{appendix1}.

 .

Since we have two baryons in the final state which are different
for each reaction, the different modification of the cross
section due to the final state interaction (FSI) of the $\Lambda
N$ or $pp$ subsystem can be relevant. We have taken this into
account by means of the ordinary factor, equivalent to the
inverse of the Jost function\footnote{Although the Jost function should not be
used for the extraction of scattering parameters from production reactions
\cite{achot}, it still gives a for our case sufficiently accurate parametrization for the
effect of the final state interaction \cite{sibham}.}. For the $\Lambda p$ interaction we
use \cite{Hinterberger:2004ra} for the factor that multiplies
$|T|^2$ in the phase space
\be
|C_{FSI}|^2=\frac{q^2+\beta^2}{q^2+\alpha^2}
\label{eq:CFSI}
\ee
\noindent
with 
\be
\alpha=(1-\sqrt{1-2r/a})/r, \beta=(1+\sqrt{1-2r/a})/r
\label{eq:alphabeta}
\ee
and $q$ the $\Lambda$ momentum in the $p\Lambda$ rest frame.
In Eq.~(\ref{eq:alphabeta}), $a$ and $r$ are the scattering
length and effective range of the $s-$wave $\Lambda p$ amplitude
in $S=1$ which is the state we have. In the calculations we use
$a=(-1.4\pm0.5)\textrm{ fm}$ and $r=(4\pm 1)\textrm{ fm}$ and we
shall estimate the uncertainties induced from these errors.

For the $pp$ final state interaction in the $pp\to p K^+ K^- p$
reaction
we use the same expression but with $a=-7.8\textrm{ fm}$
(which already accounts for interference
 with the Coulomb force \cite{bhaduri}) 
 and 
$r=2.79\textrm{ fm}$ with significant smaller errors than in the
$\Lambda p$ case.
This simple prescription is accurate to better than $10$\% in the region of
energies relevant here
 \cite{Sibirtsev:1999ka} compared to more elaborate formulas as, e.g., the one given
 in Ref. \cite{coulomb} and is therefore sufficiently accurate 
 for our purposes.

\section{Results}
\label{sec:results}

\begin{figure}
\begin{center}
\includegraphics[width=0.4\textwidth,angle=0]{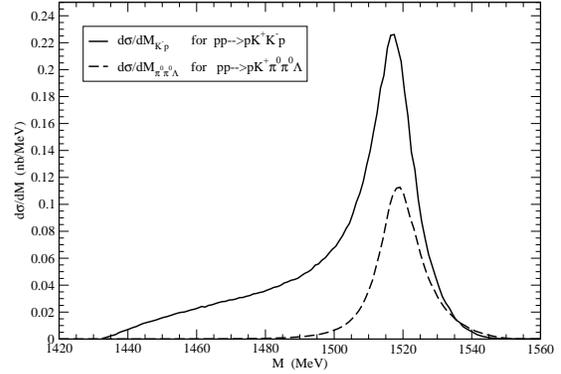}
\end{center}
\caption{Solid line:  $K^-p$ invariant mass distribution for
the  $pp\to pK^+K^-p$ reaction. Dashed line: $\pi^0\pi^0\Lambda$ invariant mass distribution
of the 
$pp\to pK^+\pi^0\pi^0\Lambda$ reaction. 
Incident proton momentum:
 $3.7\textrm{ GeV}$.}
\label{fig:MInv}
\end{figure}

First we show 
in Fig.~\ref{fig:MInv} the results for the
$\pi^0\pi^0\Lambda$ invariant mass distribution
of the 
$pp\to pK^+\pi^0\pi^0\Lambda$ reaction and the 
  $K^-p$ mass distribution for the $pp\to pK^+K^-p$
  (dashed and solid lines respectively)
for an incident proton momentum of $3.7\textrm{ GeV}$ 
which is the
highest available at COSY. This is also the nominal energy for
protons of the developing CSR facility at Lanzhou (China). 
The different shapes below the peak
for the two reactions can be understood from the presence of the
$\Sigma^*$ propagator in the $pp\to pK^+\pi^0\pi^0\Lambda$ 
reaction, since only its quickly decreasing tail enters in
the evaluation of the matrix element.

Although the production rates with this model are of the order of
experimental cross sections measured in the $pp\to pp K^+K^-$
reaction \cite{Zychor:2005sj}, other mechanisms can be at play.
In addition, we did not include the initial state interaction
that is expected to lead to a sizable reduction of the cross
section \cite{isipaper}, but would not modify the ratio of cross
sections. Because of this we focus on the shape
and relative magnitude  of our results that are model independent
as explained above. In an analogous way to what was studied in
Ref.~\cite{we} for the photoproduction case, we propose to
measure the ratio between the $\pi^0\pi^0\Lambda$ and $K^-p$ mass
distributions of the  $pp\to pK^+\pi^0\pi^0\Lambda$ and $pp\to
pK^+K^-p$ reactions respectively

\be
R\equiv\frac{d\sigma_{p p\to pK^+\pi^0\pi^0\Lambda}
/dM_{\pi^0\pi^0\Lambda}}
{d\sigma_{p p\to p K^+K^-p}
/dM_{K^-p}}
\ee

Up to phase space, FSI and known numerical factors, the ratio $R$ is
proportional to $(|T_{\pi\Sigma^*\to\pi\Sigma^*}|
/|T_{\pi\Sigma^*\to\bar{K}N}|)^2$ which, at the $\Ls$ peak position, is
$(g_1/g_3)^2=2.8$ (see Eq.~(\ref{eq:Tcoup})).  With the full model, the value
obtained for $R$ at the peak position is 
\begin{equation}
R\sim 0.5\pm 0.2 \,.
\end{equation} 
The uncertainty
given comes from the errors of $a$ and $r$ in Eq.~(\ref{eq:alphabeta}) when
considering the FSI of the final baryons, which is the largest source of
uncertainty in our model.  Had we not considered the FSI we would have
obtained $R\sim 0.9$. Actually the effect of the FSI is to increase by a
factor of $\sim3.6$ the $K^- p$ mass distribution and a factor $\sim 2$ the
$\pi^0\pi^0\Lambda$ mass distribution at the peak in the $p p\to p K^+K^-p$
and $p p\to pK^+\pi^0\pi^0\Lambda$ reactions respectively.  These factors
emerging from the final state interactions are strongly dependent on the beam
energy.

An experimental measure of this ratio $R$ would provide a good
test of the unitarized coupled channel model since the value of
$g_1$ and $g_3$ are non-trivial genuine predictions of this
theory.
Note that the close connection between the nature of a resonance
and its effective couplings to the decay channels was derived
already in Ref. \cite{wein} for bound states and extended to
inelastic resonances in Ref. \cite{baru}.

\section{Summary}

We have studied some of the consequences and predictions of a unitarized
coupled channel approach to the $\Lambda(1520)$ resonance through the $pp\to
pK^+K^-p$ and $pp\to pK^+\pi^0\pi^0\Lambda$ reactions.  The model relies upon
a coupled channel formalism implementing unitarity through the Bethe-Salpeter
equation by means of which the resonance structure appears naturally. One of
the genuine predictions of this theory are the values of the effective
couplings of the $\Ls$ resonance to the relevant channels. We propose that the
ratio between the $\pi^0\pi^0\Lambda$ and $K^-p$ mass distributions for
the $pp\to pK^+\pi^0\pi^0\Lambda$ and $pp\to pK^+K^-p$ reactions respectively,
can provide a test of the ratio between the coupling of the
$\Lambda(1520)$ resonance  to the $\pi\Sigma(1385)$
and to $\bar{K}N$ channels.
Such a ratio would help to unravel the nature of the
$\Lambda(1520)$ and its coupling to meson-baryon components.

\section*{Acknowledgments}
We thank A. Sibirtsev for useful comments.
This work is partly supported by DGICYT contract number BFM2003-00856,
and the E.U. EURIDICE network contract no. HPRN-CT-2002-00311 and
the Deutsche Forschungsgemeinschaft  through funds provided to the SFB/TR 16
``Subnuclear Structure of Matter''. 
This research is part of the EU Integrated Infrastructure Initiative
Hadron Physics Project under contract number RII3-CT-2004-506078.


\appendix

\section{Five- and four-body phase space}
\label{appendix1}

Here we explain in detail the steps followed to simplify
the integrals of the phase space in the evaluation of the cross
section for the 
\be
p(q_1)p(q_2)\to p(p_5)K^+(p_4)\pi^0(p_1)\pi^0(p_2)\Lambda(p_3)
\ee
reaction (see Fig.~\ref{fig:diagrampipiL} for the detailed
definition of the momenta) with masses $M_1$, $M_2$, $m_5$,
$m_4$, $m_1$, $m_2$ and $m_3$ respectively.

The cross section for the process is given by:

\ba
\sigma&=&\frac{2SM_1M_2}{\sqrt{s^2-2s(M_1^2+M_2^2)
    +(M_1^2-M_2^2)^2}} \nonumber\\
 &\times&\int \frac{d^3p_1}{(2\pi)^3} \int\frac{d^3p_2}{(2\pi)^3} 
 \int\frac{d^3p_3}{(2\pi)^3}\int \frac{d^3p_4}{(2\pi)^3} 
 \int\frac{d^3p_5}{(2\pi)^3}  \nonumber\\
&\times& \frac{1}{2\omega_1} \frac{1}{2\omega_2}
 \frac{m_3}{\omega_3}\frac{1}{2\omega_4}\frac{m_5}{\omega_5}
 \nonumber\\
&\times&
 (2\pi)^4\delta^4(q_1+q_2-\sum p_i)
 |Amp|^2,
 \label{eq:crosspipiL}
\ea
\noindent
where $S=1/2$ for the symmetry of $\pi^0\pi^0$ and
$p^\mu_i=(\omega_i,\vec p_i)$.

The evaluation of the $\delta$ of energy conservation for so
many particles can be a difficult task. However it can be
extremely simplified by using the following procedure.

The quantity 
\ba
A&\equiv&\int\frac{d^3p_4}{(2\pi)^3}\int\frac{d^3p_3}{(2\pi)^3}
\frac{1}{\omega_3\omega_4}\nonumber\\
&\times&
(2\pi)^4\delta^4(q_1+q_2-\sum p_i)
 |Amp|^2\nonumber\\
&=&\int\frac{d^3p_4}{(2\pi)^3}\frac{1}{\omega_3\omega_4}
\nonumber\\
&\times&
(2\pi)\delta(q_1^0+q_2^0-\sum \omega_i)
 |Amp|^2
\ea
is Lorentz invariant, hence we can evaluate it in any desired
frame. By convenience we choose a frame where 
$\vec{p_3}+\vec{p_4}=0$. In this frame the $\delta$ function of
energy conservation is simple since 
$M_{34}^2=(p_3+p_4)^2=(q_1+q_2-p_1-p_2-p_5)^2
=(q_1^0+q_2^0-p_1^0-p_2^0-p_5^0)^2$
and hence
$\delta(q_1^0+q_2^0-p_1^0-p_2^0-\omega_3-\omega_4-p_5^0)
=\delta(M_{34}-\sqrt{p_4^2+m_3^2}-\sqrt{p_4^2+m_4^2})$
 is only a function of $|\vec{p_4}|$ and
  hence can be use to evaluate
the $|\vec{p_4}|$ integral, which gives
\be
A=\int d\cos\theta'_4 d\phi'_4|\vec{p'_4}|
\frac{1}{(2\pi)^2}\frac{|\vec{p'_4}|}{M_{34}} |Amp|^2
\Theta(M_{34}-m_3-m_4)
\ee
\noindent
where $|\vec{p'_4}|=\lambda^{1/2}
(M^2_{34},m_3^2,m_4^2)/2M_{34}$ is the momentum in the 
$\vec{p_3}+\vec{p_4}=0$ frame.

The final expression for the cross section is 
\ba
\sigma&=&\frac{SM_1M_2m_3m_5}{4(2\pi)^{11}\sqrt{s^2-2s(M_1^2+M_2^2)
    +(M_1^2-M_2^2)^2}}\nonumber\\
&\times&
 \int d^3p_1 \int d^3p_2 \int d^3p_5
 \int d\cos\theta'_4 \int d\phi'_4  \nonumber\\
&\times&
 \frac{|\vec{p'_4}|}
 {\omega_1\omega_2\omega_5M_{34}}\Theta(M_{34}-m_3-m_4)
 |Amp|^2.
\ea
When evaluating the amplitude all the momenta have to be
evaluated in the overall CM frame, hence one has to boost back
the $\vec{p'_4}$ momentum to this system with the following
expression:
\be
\vec{p_4}=\vec{p'_4}+\left[\left(\frac{P^0}{M_{34}}-1\right)
\frac{\vec{p'_4}\cdot\vec{P}}{|\vec{P}|^2}
+\frac{p_4^{'0}}{M_{34}}\right]\vec{P}.
\label{eq:boostp4}
\ee
with $\vec{P}=-\vec{p_1}-\vec{p_2}-\vec{p_5}$ and 
$P^0=\sqrt{s}-p_1^0-p_2^0-p_5^0$.\\

The four-body phase space for the 
\be
p(q_1)p(q_2)\to p(p_2)K^+(p_4)K^-(p_3)p(p_1)
\ee
 reaction can be
evaluated in an analogous way to the previous one and gives
\ba
\sigma&=&\frac{SM_1M_2m_1m_2}{2(2\pi)^{8}
\sqrt{s^2-2s(M_1^2+M_2^2)+(M_1^2-M_2^2)^2}}\nonumber\\
&\times&
 \int d^3p_1 \int d^3p_2 
 \int d\cos\theta'_4 \int d\phi'_4  
 \frac{|\vec{p'_4}|}
 {\omega_1\omega_2M_{34}}\nonumber\\
&\times&
 \Theta(M_{34}-m_3-m_4)
 |Amp|^2,
 \label{eq:crossKp}
\ea
where $S=1/2$ since the final protons are identical particles.
The boost of $p'_4$ to the overall CM frame is done with
Eq.~(\ref{eq:boostp4}) but now 
$\vec{P}=-\vec{p_1}-\vec{p_2}$ and 
$P^0=\sqrt{s}-p_1^0-p_2^0$.

\section{Alternative derivation of the amplitudes}
\label{app:ampl}

Here we explain in detail the analytic evaluation of the amplitudes for the
$pp\to pK^+\pi^0\pi^0\Lambda$ and $pp\to pK^+K^-p$ reactions within the method
of irreducible tensors, which is more familiar to many readers than the
technique described in the main text.

 Let us start by determining the allowed quantum
numbers of the initial and final states, which are independent of the internal
mechanisms, which will be of help in simplifying the evaluation of the
amplitudes later on. Let us consider the general process $pp\to K^+ p\Ls$.
Since the reaction is calculated at energies close to threshold, the only
final total angular momentum allowed is $L=0$. Hence the final $J^P$
possibilities are $1^+$ or $2^+$, since $J^P(K^+)=0^-$, $J^P(p)=1/2^+$ and
$J^P(\Ls)=3/2^-$. The spin of the initial $pp$ pair can be $0$ or $1$ while
$L$ has to be even for parity reasons. But, since the initial protons are
identical fermions, $L+S+I$ has to be odd. And, since, $I=1$, then $L+S$ has
to be even. Hence the only possibility to match final $1^+$ or $2^+$ is that
the initial protons are in $L=2$ and $S=0$. In summary, the initial $pp$ state
has the following quantum numbers: $L=2$, $S=0$, $P=+$, independent of the
internal dynamics and the $\Ls$ decay products.

Let us consider now the quantum numbers of the $p\Lambda$ system
in the 
$pp\to pK^+(\Ls)\to  pK^+(\pi^0\Sigma^*)\to
pK^+(\pi^0\pi^0\Lambda)$ reaction.
Considering the parity of the final particles and the fact that
the $\Ls$ decay into $\pi\Sigma^*$ is in s-wave and the decay of
the $\Sigma^*$ into $\pi\Lambda$ is in p-wave, the angular
momentum of the $p\Lambda$ system has to be even to match the global
parity $+$. Given the typical momenta of the final $\Lambda$ at
the energies of concern in the present work,
 the only possibility is
$L(p\Lambda)=0$. On the other hand the spin of the $p\Lambda$
system has to be $S(p\Lambda)=1$ in order to give total
 $J=2$ when
combined to the p-wave of the $\Sigma^*\to\pi\Lambda$ decay in line with the
formalism as described in the main text.

Regarding the $pp\to pK^+(\Ls)\to pK^+(K^-p)$ reaction, the kinematics of the
$pp$ final state is that the proton coming from the $\Lambda(1520)$ decay has
about $240\textrm{ MeV/c}$ momenta since the $\Ls$ is produced essentially at
rest, and the other proton is also basically at rest in the total CM frame.
This gives the protons a kinetic energy of about $8\textrm{ MeV}$ in their CM
frame where only relative S-waves are relevant. Hence, the final $pp$ system
only can be in $L(pp)=0$ at the energies considered.  On the other hand the
spin is $S(pp)=0$ since $L+S+I$ has to be odd.

Let us evaluate the amplitude for the $pp\to
pK^+\pi^0\pi^0\Lambda$ reaction of Fig.~\ref{fig:diagrampipiL}.
The amplitude for the vertex $pK^+K^-p$ is given by
 \cite{Oset:1997it}
\be
-it=-i\frac{1}{2f^2}(p_4^0-k^0).
\ee
The unitarized $\bar{K}N\to\pi\Sigma^*$ transition has the
following form \cite{Sarkar:2005ap}
\ba
-it_{K^-p\rw\pi^0\Sigma^{*0}}&=&-i
\frac{1}{\sqrt{2}}\frac{(-1)}{\sqrt{3}}
T_{\bk N\rw\pi\Sgs}\ 
{\cal C}(\frac{1}{2}\ 2\
\frac{3}{2};-m,M+m)\nonumber\\
&\times&Y_{2,-m-M}(\hat{k})(-1)^{M+m}\sqrt{4\pi},
\ea
\noindent
where ${\cal C}$ are Clebsch-Gordan coefficients.
For the decay of the $\Sigma^*$ into $\pi^0\Lambda$, the vertex
is \cite{Oset:2000eg}
\be
-it=-\frac{f_{\Sigma^*\pi\Lambda}}{m_\pi}
\langle\frac{1}{2} m''|\vec{S}\cdot\vec{p_2}|\frac{3}{2} M\rangle.
\ee
\noindent
where
 $f_{\Sgs\pi\Lambda}/m_\pi=9.61\times10^{-3}\textrm{ MeV}^{-1}$
and $\vec{S}$ is the total spin $3/2$ to $1/2$ transition
operator given by:
\begin{equation}
\vec{S}\cdot\vec{p} =\left(\begin{array}{cccc} 
-\frac{p_x+ip_y}{\sqrt{2}}&\sqrt{\frac{2}{3}}p_z
& \frac{p_x-ip_y}{\sqrt{6}} & 0 \\
0&-\frac{p_x+ip_y }{\sqrt{6}}
& \sqrt{\frac{2}{3}}p_z&\frac{p_x-ip_y}{\sqrt{2}}
\end{array}
\right)
\end{equation}

Hence, the full amplitude is given by
\ba
t&=&i\frac{1}{2f^2}(p_4^0-k^0)\frac{1}{k^2-m_K^2}
\frac{1}
{\sqrt{s_{\Sgs}}-M_{\Sgs}+i\Gamma_{\Sgs}(\sqrt{s_{\Sgs}})/2}
\nonumber\\
&\times& \frac{f_{\Sgs\pi\Lambda}}{m_\pi}\frac{1}{\sqrt{6}} 
T_{\bk N\rw\pi\Sgs}\ {\cal C}(\frac{1}{2}\ 2\
\frac{3}{2};-m,M+m)\nonumber\\
&\times&Y_{2,-m-M}(\hat{k})(-1)^{M+m}\sqrt{4\pi}\,
\langle\frac{1}{2} m''|\vec{S}\cdot\vec{p_2}
|\frac{3}{2} M\rangle \nonumber \\
&\times&\delta_{m',m}\delta_{M,-m},
\label{eq:fullt1}
\ea
\noindent
where $s_{\Sgs}=(p_2+p_3)^2$.

Since the momentum of the exchanged kaon is approximately equal
to the incoming proton momentum, we can do the following
approximation:
\be
Y_{2,-m-M}(\hat{k})\simeq
Y_{2,-m-M}(\hat{u_z})=\sqrt{\frac{5}{4\pi}}.
\ee

On the other hand, the spin of the initial 
$pp$ system is 
$S(pp)=0$ (antisymmetric)
and the spin of the final $p\Lambda$ system is $S(p\Lambda)=1$
(symmetric), as discussed above. Therefore, 
the initial spin wave function has to be
antisymmetrized and the final $p\Lambda$ spin wave function has
to be symmetrized.
Let us consider only the spin dependent part of 
Eq.~(\ref{eq:fullt1}).
The antisymmetrization of the initial spin gives

\ba
&&
\frac{1}{\sqrt{2}}\left[
{\cal C}(\frac{1}{2}\ 2\ \frac{3}{2};-\frac{1}{2}\ 0)
\langle\frac{1}{2} m''|\vec{S}\cdot\vec{p_2}
|\frac{3}{2} -\frac{1}{2}\rangle
\delta_{m',+1/2} \right.\nonumber \\
&&\ \ \ \ -\left.{\cal C}(\frac{1}{2}\ 2\ \frac{3}{2};
+\frac{1}{2}\ 0)
\langle\frac{1}{2} m''|\vec{S}\cdot\vec{p_2}
|\frac{3}{2} +\frac{1}{2}\rangle
\delta_{m',-1/2}\right]\nonumber\\
&=&\frac{1}{\sqrt{2}}\sqrt{\frac{2}{5}}
\left[\langle\frac{1}{2} m''|\vec{S}\cdot\vec{p_2}
|\frac{3}{2} -\frac{1}{2}\rangle
\delta_{m',+1/2}\right.\nonumber \\
&+&\left.\langle\frac{1}{2} m''|\vec{S}\cdot\vec{p_2}
|\frac{3}{2} +\frac{1}{2}\rangle
\delta_{m',-1/2}\right].
\ea

Now we have to symmetrize the latter expression for the final
$p\Lambda$ spin.
For $S_z(p\Lambda)=0$ gives
\ba
&&\frac{1}{\sqrt{2}}\frac{1}{\sqrt{2}}\sqrt{\frac{2}{5}}
\left[\langle\frac{1}{2} -\frac{1}{2}|\vec{S}\cdot\vec{p_2}
|\frac{3}{2} -\frac{1}{2}
\rangle + 0\right.\nonumber \\
&&\ \ \ \ \ \ \ \ \ \ \ \ \ 
+\left.0+\langle \frac{1}{2} +\frac{1}{2}|\vec{S}\cdot\vec{p_2}
|\frac{3}{2} +\frac{1}{2}\rangle
\right] \nonumber \\
&&=
\sqrt{\frac{2}{5}}\sqrt{\frac{2}{3}} p_{2z},
\ea

for $S_z(p\Lambda)=-1$:
\ba
&&\frac{1}{\sqrt{2}}\sqrt{\frac{2}{5}}
\left[0+ 
\langle\frac{1}{2} -\frac{1}{2}|\vec{S}\cdot\vec{p_2}
|\frac{3}{2} +\frac{1}{2}\rangle\right]
 \nonumber \\
&&=-\frac{1}{\sqrt{5}}\frac{1}{\sqrt{6}} (p_{2x}+ip_{2y}),
\ea

and for $S_z(p\Lambda)=+1$:
\ba
&&\frac{1}{\sqrt{2}}\sqrt{\frac{2}{5}}
\left[ 
\langle\frac{1}{2} +\frac{1}{2}|\vec{S}\cdot\vec{p_2}
|\frac{3}{2} -\frac{1}{2}\rangle+0\right]
 \nonumber \\
&&=\frac{1}{\sqrt{5}}\frac{1}{\sqrt{6}}
(p_{2x}-ip_{2y}).
\ea

Hence, the final expression for the amplitude is

\ba
t&=&-i\frac{1}{3}\frac{1}{2f^2}(p_4^0-k^0)
\frac{1}{k^2-m_k^2}\nonumber\\
&\times&
\frac{1}
{\sqrt{s_{\Sgs}}-M_{\Sgs}+i\Gamma_{\Sgs}(\sqrt{s_{\Sgs}})/2}
\frac{f_{\Sgs\pi\Lambda}}{m_\pi}  \nonumber \\
&\times&T_{\bar{K}N \to\pi\Sigma^*}
\left\{\begin{array}{l}{\sqrt{2}p'_{2z}~~~~~~~~~~~~~S_z=0}\\
{\frac{1}{2}(p'_{2x}-ip'_{2y})~~~~S_z=+1}\\
{-\frac{1}{2}(p'_{2x}+ip'_{2y})~~S_z=-1}
\end{array}\right\}
\label{eq:apptpipiL}
\ea
\noindent
and the symmetrization of the identical pions has to be
considering by adding, for each $S_z$, the amplitude
changing $p_1\leftrightarrow p_2$.

For the evaluation of the $pp\to pK^+K^-p$ amplitude
we need the unitarized $K^-p\to K^-p$ amplitude which is given by
\ba
-it_{K^-p\rw K^-p}&=&-i\frac{1}{2}T_{\bk N\rw\bk N}
\sum_M {\cal C}(\frac{1}{2}\ 2\ \frac{3}{2};-m,M+m)\nonumber\\
&\times&
Y_{2,-m-M}(\hat{k})
{\cal C}(\frac{1}{2}\ 2\ \frac{3}{2};m'',M-m'')\nonumber\\
&\times&
Y^*_{2,m''-M}(\hat{p_3})(-1)^{m''+m}4\pi.
\label{eq:34}
\ea
The rest of the procedure is analogous to the previous case
but now the final $pp$ spin is $S(pp)=0$, as discussed above,
which is antisymmetric and hence we have now 
to antisymmetrize the
final $pp$ spin wave function. 

The final expression of the amplitude is given by
\be
t(p_1,p_2)=\frac{1}{2f^2}(p_4^0-k^0)
\frac{1}{k^2-m_k^2}
\frac{1}{2}(3\cos^2\theta_3-1) 
T_{\bar{K}N \to\bar{K}N }
\label{eq:apptKN}
\ee
where the symmetry of the final $pp$ space wave function has to
be considered by adding to the amplitude the same expression
changing $p_1\leftrightarrow p_2$.



\begin{thebibliography}{99}


\bibitem{Kaiser:1996js}
  N.~Kaiser, T.~Waas and W.~Weise,
  Nucl.\ Phys.\ A {\bf 612} (1997) 297.
  
\bibitem{Oset:1997it}
E.~Oset and A.~Ramos,
Nucl. Phys. A {\bf 635} (1998) 99.

\bibitem{Meissner:1999vr}
U.-G.~Mei{\ss}ner and J.~A.~Oller,
Nucl.\ Phys.\ A {\bf 673} (2000) 311
[arXiv:nucl-th/9912026].

\bibitem{Oller:2000fj}
J.~A.~Oller and U.-G.~Mei{\ss}ner,
Phys.\ Lett.\ B {\bf 500} (2001) 263.

\bibitem{Garcia-Recio:2002td}
  C.~Garcia-Recio, J.~Nieves, E.~Ruiz Arriola and M.~J.~Vicente Vacas,
  Phys.\ Rev.\ D {\bf 67} (2003) 076009.

\bibitem{Hyodo:2003jw}
T.~Hyodo, A.~Hosaka, E.~Oset, A.~Ramos and M.~J.~Vicente Vacas,
Phys.\ Rev.\ C {\bf 68} (2003) 065203.






\bibitem{Oller:2000ma}
J.~A.~Oller, E.~Oset and A.~Ramos,
Prog.\ Part.\ Nucl.\ Phys.\  {\bf 45} (2000) 157.

\bibitem{Jido:2003cb}
D.~Jido, J.~A.~Oller, E.~Oset, A.~Ramos and U.-G.~Mei{\ss}ner,
Nucl.\ Phys.\ A {\bf 725} (2003) 181.


\bibitem{Garcia-Recio:2003ks}
  C.~Garcia-Recio, M.~F.~M.~Lutz and J.~Nieves,
  Phys.\ Lett.\ B {\bf 582} (2004) 49.

\bibitem{Prakhov:2004an}
  S.~Prakhov {\it et al.}  [Crystall Ball Collaboration],
  Phys.\ Rev.\ C {\bf 70} (2004) 034605.

\bibitem{Thomas:1973uh}
  D.~W.~Thomas, A.~Engler, H.~E.~Fisk and R.~W.~Kraemer,
  Nucl.\ Phys.\ B {\bf 56} (1973) 15.

\bibitem{Magas:2005vu}
V.~K.~Magas, E.~Oset and A.~Ramos,
Phys.\ Rev.\ Lett.\  {\bf 95} (2005) 052301.

\bibitem{lutz}
E.~E.~Kolomeitsev and M.~F.~M.~Lutz,
Phys.\ Lett.\ B {\bf 585} (2004) 243.

\bibitem{Sarkar:2004jh}
S.~Sarkar, E.~Oset and M.~J.~Vicente Vacas,
Nucl.\ Phys.\ A {\bf 750} (2005) 294.

\bibitem{nakano}  
T. Nakano, talk at the Pentaquark04 meeting
http://www.rcnp.osaka-u.ac.jp/$\sim$ penta04/

\bibitem{Hicks:2005gp}
K.~H.~Hicks,
Prog.\ Part.\ Nucl.\ Phys.\  {\bf 55} (2005) 647.



\bibitem{Eidelman:2004wy}
  S.~Eidelman {\it et al.}  [Particle Data Group],
  Phys.\ Lett.\ B {\bf 592}, 1 (2004).

\bibitem{Mast:1973gb}
T.~S.~Mast, M.~Alston-Garnjost, R.~O.~Bangerter, A.~Barbaro-Galtieri, F.~T.~Solmitz and R.~D.~Tripp,
Phys.\ Rev.\ D {\bf 7} (1973) 5.





\bibitem{Prakhov:2004ri}
S.~Prakhov {\it et al.},
Phys.\ Rev.\ C {\bf 69} (2004) 042202.


\bibitem{Sarkar:2005ap}
S.~Sarkar, E.~Oset and M.~J.~Vicente Vacas,
Phys.\ Rev.\ C {\bf 72} (2005) 015206.


\bibitem{we}
L.~Roca, Sourav~Sarkar, V.~K.~Magas and E.~Oset, {\it Submitted to
Phys.\ Rev.\ C}
; 
S.~Sarkar, L.~Roca, E.~Oset, V.~K.~Magas and M.~J.~V.~Vacas,
arXiv:nucl-th/0511062.

\bibitem{ollerND}
J.~A.~Oller and E.~Oset,
 Phys.\ Rev.\ D {\bf 60} (1999) 074023.

\bibitem{manohar}
E.~Jenkins and A.~V.~Manohar,
Phys.\ Lett.\ B {\bf 259} (1991) 353.

\bibitem{meier}
R.~Maier, {\it private communication}

\bibitem{Zychor:2005sj}
I.~Zychor {\it et al.},
Phys.\ Rev.\ Lett.\  {\bf 96} (2006) 012002
[arXiv:nucl-ex/0506014].

\bibitem{report} C. Hanhart, Phys. Rept. {\bf 397}, 155 (2004).

\bibitem{achot}
A.~Gasparyan, J.~Haidenbauer and C.~Hanhart,
  Phys.\ Rev.\ C {\bf 72} (2005) 034006.


\bibitem{sibham}
  A.~Sibirtsev, J.~Haidenbauer, H.~W.~Hammer and S.~Krewald,
  arXiv:nucl-th/0512059.



\bibitem{Hinterberger:2004ra}
F.~Hinterberger and A.~Sibirtsev,
Eur.\ Phys.\ J.\ A {\bf 21} (2004) 313.

\bibitem{bhaduri}
M.~A.~Preston and R.~K.~Bhaduri,  
{\it Structure of the nucleus}, Addison Wesley 1975, pags. 39-51.
 

\bibitem{Sibirtsev:1999ka}
A.~Sibirtsev and W.~Cassing,
Eur.\ Phys.\ J.\ A {\bf 7} (2000) 407.

\bibitem{coulomb}
B.~L.~Druzhinin, A.~E.~Kudryavtsev and V.~E.~Tarasov,
  Z.\ Phys.\ A {\bf 359}, 205 (1997).

\bibitem{isipaper} C. Hanhart and K. Nakayama,
Phys.\ Lett.\ B {\bf 454} (1999) 176. 

\bibitem{wein}
 S.~Weinberg,
  Phys.\ Rev.\  {\bf 130} (1963) 776.
  

\bibitem{baru}
 V.~Baru et al.,
  Phys.\ Lett.\ B {\bf 586} (2004) 53
  [arXiv:hep-ph/0308129].

\bibitem{Oset:2000eg}
E.~Oset and A.~Ramos,
Nucl.\ Phys.\ A {\bf 679} (2001) 616.



\end{thebibliography}
\end{document}